# A map of Digital Humanities research across bibliographic data sources

Gianmarco Spinaci, gianmarco.spinaci@studio.unibo.it, https://orcid.org/0000-0002-3504-3241
Villa I Tatti, The Harvard University Center for Italian Renaissance Studies, Harvard University, Florence, Italy

Giovanni Colavizza, g.colavizza@uva.nl, https://orcid.org/0000-0002-9806-084X
Institute for Logic, Language and Computation, University of Amsterdam, Amsterdam, The Netherlands

Silvio Peroni, silvio.peroni@unibo.it, https://orcid.org/0000-0003-0530-4305
Research Centre for Open Scholarly Metadata, Department of Classical Philology and Italian Studies, University of Bologna, Bologna, Italy
Digital Humanities Advanced Research Centre, Department of Classical Philology and Italian Studies, University of Bologna, Bologna, Italy

## Abstract

**Purpose.** This study presents the results of an experiment we performed to measure the coverage of Digital Humanities (DH) publications in mainstream open and proprietary bibliographic data sources, by further highlighting the relations among DH and other disciplines.
**Methodology.** We created a list of DH journals based on manual curation and bibliometric data. We used that list to identify DH publications in the bibliographic data sources under consideration. We used the ERIH-PLUS list of journals to identify Social Sciences and Humanities (SSH) publications. We analysed the citation links they included to understand the relationship between DH publications and SSH and non-SSH fields.
**Findings.** Crossref emerges as the database containing the highest number of DH publications. Citations from and to DH publications show strong connections between DH and research in Computer Science, Linguistics, Psychology, and Pedagogical & Educational Research. Computer Science is responsible for a large part of incoming and outgoing citations to and from DH research, which suggests a reciprocal interest between the two disciplines.
**Value.** This is the first bibliometric study of DH research involving several bibliographic data sources, including open and proprietary databases.
**Research limitations.** The list of DH journals we created might be only partially representative of broader DH research. In addition, some DH publications could have been cut off from the study since we did not consider books and other publications published in proceedings of DH conferences and workshops. Finally, we used a specific time coverage (2000-2018) that could have prevented the inclusion of additional DH publications.

## 1. Introduction

Despite many attempts to answer the open-ended question "What are the Digital Humanities?" (McCarty, 2008; Scholes et al., 2008; Svensson, 2009; Weingart et al., 2017; Roth, 2019; Münster et al., 2020, Earhart et al., 2020), currently there is not a shared answer. On the one hand, this allows one to include a broad spectrum of approaches and methodologies under a big Digital

Humanities tent. On the other hand, it makes it challenging to grasp the field's internal and external organisation and see it recognised as a discipline on its own merits (Piotrowski, 2020). From a bibliometric point of view, without a clear field delineation of a starting set of Digital Humanities publications, it is challenging to pursue subsequent and more specific analyses. At the same time, empirical studies of how the Digital Humanities are intellectually organised would offer an important complement to the debate (Poole, 2017).

The term Digital Humanities (DH) was introduced only a few decades ago (Schreibman et al., 2004). Recent research aimed at unveiling their intellectual organisation at the use of bibliometric methods. Recent work applying topic modelling (a text mining technique) to a set of DH journals found that the DH are highly interdisciplinary and strongly related to computational linguistics and information science (Luhmann et al., 2021). Nevertheless, because of the vast spectrum of publications and lack of a distinct class identifying DH publication in mainstream bibliographic data sources, those analyses are few and still leave several unanswered questions. The starting point of bibliographic studies of the DH is thus an open challenge.

In this study, we aim to measure the coverage of DH publications in mainstream open and proprietary bibliographic data sources, by highlighting the relation between DH and other disciplines, i.e., which disciplines are cited by DH research and which ones cite DH studies. To achieve this result, we introduce a methodology to identify DH publications and their citations. In the context of this article, a citation is defined as a conceptual directional link from a citing entity to a cited entity, created through particular textual devices included in the content of the citing entity (e.g. a bibliographic reference denoted by an in-text reference pointer such as "[3]" or "(Doe et al., 2021)") for acknowledging others' work for some reason, such as to reference others' ideas, arguments, software, and methodologies that are discussed, used, or criticized in the citing entity. To identify DH publications, we built a list of DH journals created via a combination of bibliometric data and manual curation. We considered only journal articles indexed in the sources focused on publishing exclusively or primarily articles that could be considered part of the DH. The data sources considered were *Crossref* (Hendricks et al., 2020) complemented with the *OpenCitations Index of Crossref open DOI-to-DOI citations* (COCI) (Heibi et al., 2019) for gathering citation data, *Dimensions* (Herzog et al., 2020), *Microsoft Academic Graph* (Wang et al., 2020), *Scopus* (Baas et al., 2020), and *Web of Science*, including its *Arts & Humanities Citation Index* (A&HCI) (Birkle et al., 2020). These data sources presented different coverage in terms of the number of DH publications they included due to their internal policies for selecting relevant publications and venues to store in their catalogue. In order to understand how the DH relate to other disciplines in the Arts & Humanities (AH) and Social Sciences (SS), we relied on the *European Reference Index of the Humanities and Social Sciences* (ERIH PLUS, https://dbh.nsd.uib.no/publiseringskanaler/erihplus/). Instead, we used Web of Science's Research Areas (https://images.webofknowledge.com/images/help/WOS/hp_research_areas_easca.html) to map the relation between DH research and other non-SSH disciplines.

## 2. Related Work

Bibliographic data sources are a crucial asset in bibliometric research, for example, allowing the creation and study of maps of science (Boyack et al., 2005; Boyack et al., 2009). The use of maps of science as a technique to study the cognitive structure of Humanities disciplines has been in use at least since the 1980s, slowly yet steadily expanding in more recent years (Franssen et al.,

2019). The Arts and Humanities Citation Index (A&HCI), now part of Web of Science, has often been instrumental in such studies of the Humanities (Small et al., 1985; Hicks et al., 2009). As an important example, Leydesdorff et al. (2010) applied techniques previously used for mapping the journal structure of the sciences and social sciences to the A&HCI, using the ERIH PLUS list of disciplines to identify Arts & Humanities (AH) research.

However, bibliographic data sources are known to contain limitations, in particular for AH research. First, the identification of AH works is not trivial. For example, it is sometimes difficult to decouple the humanities from the social sciences because they are often mixed (Larivière et al., 2006). Archambault et al. (2010) identified three further limitations: the low coverage of Humanities publications in bibliographic data sources, their publication ageing rate (i.e., how much publications stay mainstream), and their low citation rate. Thus, not surprisingly, there is still a sense of mistrust of bibliometrics when applied to the Humanities (Kosmopoulos et al., 2007; Bornmann et al., 2016; Hammarfelt, 2016; Sivertsen, 2016; Melchiorsen, 2019).

Recently, Visser et al. (2021) analysed the overall coverage of the Social Sciences and Humanities (SSH) in Scopus, Web of Science, Dimensions, Crossref, and Microsoft Academic Graph, considered journal articles, books, and conference proceedings. The authors noticed an overall low coverage of SSH publications when compared to STEM fields, and a significant lack of metadata in Microsoft Academic Graph. A similar result has been obtained by Singh et al. (2021), who analysed publications in Web of Science, Scopus, and Dimensions published between 2010 and 2018, showing that the percentages of SSH publications were comparatively low compared to other fields – even though Dimensions appeared to have a significantly better coverage of SSH publications.

The use of empirical bibliometric evidence to understand how a field of research is organised is of particular importance for young disciplines, such as Digital Humanities (DH), where identity is still heavily debated (Poole, 2017). During the last decade, some in the DH community proposed to see the field as a *big tent* (Pannapacker, 2011; Jockers et al. 2011), even though that view has been criticised as essentially a way to dodge the question (Terras, 2013; Piotrowski, 2020). A considerable amount of DH research is published in journals and conference proceedings. This scenario would enable, in principle, to use bibliometrics approaches to empirically find whether DH is a big tent or a well-defined field of research.

Indeed, recently several works have tackled DH from a bibliometric point of view. Such studies often have used publication venues to analyse DH research, such as in (Svensson, 2009), where the author studied the first issue of *Digital Humanities Quarterly* (DHQ) along with data from DH research centres. More recently, Sula et al. (2019) used the articles published between 1960 and 2004 in *Computers in the Humanities* (CHum) and *Literary and Linguistic Computing* (LLC) and focused on understanding the multidisciplinary aspects of DH research and the relation that may exist with other disciplines such as Computational Linguistics. Gao et al. (2017) used the same set of publications considered in (Sula et al., 2019) plus those included in DHQ to study co-citation networks of DH publications and found that DH research tended to be distributed around prominent authors and specific research areas such as Historical Literacy, Information Science, Language Modelling and Natural Language Processing, Statistics, and Text Analysis.

In a recent study, Luhmann et al. (2021) used topic modelling on a set of DH research published in *Digital Scholarship in the Humanities* (DSH), CHum, LLC, DHQ between 1990 and 2019, and compared them against a sample from other disciplines. Their main question was to understand whether DH is a field of research on its own, rather than a meta-field connecting the Humanities with Computer Science. Their findings showed that DH research presents features of both types. They also found strong topical proximity between DH, Computational Linguistics and Information Science research, which confirmed results of previous studies obtained from analysing papers published at the DH Conference 2015 (Weingart, 2014) and bibliographic metadata in Web of Science (Wang, 2018).

Using the same list of DH journals as our study (Spinaci et al., 2020), Ma et al. (2021) retrieved a sample of 252 articles from Crossref and analysed their content relying on the classification scheme described in (Sula et al., 2019). They noticed that authors of DH publications grounded on STEM research usually publish in cross-domain journals, such as the *Journal on Computing and Cultural Heritage* (JOCCH) published by the *Association for Computing Machinery* (ACM), while authors were addressing DH topics through pure Humanities methodologies publish on traditional DH venues such as DSH. This finding is important in retrospect, given the use of a selection of such traditional DH journals by several previous studies discussed above.

While these bibliometric studies have provided important evidence on the empirical intellectual and social organisation of the DH, to the best of our knowledge no systematic map of DH research exists across bibliographic data sources other than the SSH, which has an important tradition of coverage analysis. For these reasons, the coverage of these databases over the Digital Humanities is unknown.

# 3. A List of Digital Humanities Journals

In this work, which is a follow-up of a previous analysis we did (Spinaci et al., 2020), we aim at expanding on the issue of the coverage of DH publications in the main open and proprietary citation databases, also analysing the relationship between DH and SSH fields, which have not been addressed before at a large scale.

To this end, we decided to use an approach already used in the literature (Leydesdorff et al., 2010), i.e., the adoption of a list of DH journals to identify DH publications. However, the lists of DH journals currently available online at the time of this study, for example, the *Alliance of Digital Humanities Organisations* (ADHO) list (https://adho.org/publications) and the list curated by Berkeley (Berkeley, 2016), while an excellent starting point, were not deemed comprehensive enough to justify their use in our large-scale investigation since they did not include several DH journals of our knowledge that could be relevant to include in our study. For this reason, we decided to create a new list of DH journals based on manual curation and bibliometric data. In this section, we introduce the methodology we used to create such as a list of DH journals, which is organised in three sequential steps. We only acknowledged the release of the *Digital Humanities Literacy Guidebook* (Weingart et al., 2021) after our project started, so we decided to ingest it in the last step.

## 3.1. Step 1: crowdsourcing an initial set of DH journal

We identified an initial list of DH journals by involving DH experts from different geographical and linguistic areas and various disciplinary perspectives. We asked all scholars participating in the DH Conference 2019 via social media accounts and the Humanist Discussion Group (https://dhhumanist.org) to fill in a Google Form survey with the aim of crowdsourcing DH journals metadata. In the survey, we optionally required participants to indicate their job title, affiliation, and nationality. We were able to involve fourteen DH experts: five from Italy, three from the United States, two from Hungary, and one each from Canada, Spain, Australia, and Sweden.

We asked them to suggest up to ten DH journals each by specifying their basic metadata, including title, ISSN, URL, the perceived primary language for each journal, and the *DH level* of the journal, set to one of the following options:

- *Exclusively* – the journal focuses entirely on DH;
- *Significantly* – the journal publishes a high amount of DH publications, but DH is not its primary focus;
- *Marginally* – the journal publishes a relatively small amount of DH contributions;

In the resulting list of journals, we have classified them according to the answers given by the experts. In the event of disagreement (e.g., when two or more experts suggested different classifications for the same journal), we resolved it among the authors, by selecting the best fitting classification among the ones selected by the experts.

## 3.2. Step 2: enriching the DH journal list via citation links

Using the version of Web of Science (WoS) (as of the 13th week of 2018) curated and clustered according to the Leiden Algorithm (Traag el al., 2019) by the scientometricians of the CWTS (https://www.cwts.nl), we obtained the bibliographic metadata and citation data of all the publications in the DH journals identified by the experts involved in the first step, organised in clusters. For each cluster that included more than 5% of DH publications, we picked the top-5 journals by publication counts (within the cluster) and added them to the list of DH journals if they were not already identified by our experts during the first step. We selected this specific threshold because we noticed a gap in the magnitude between clusters with more than 5% of DH publications and the others. Specifically, the clusters with a percentage value less than the threshold were also fewer than 1%.

Next, each author of the present article labelled these newly identified journals independently with the DH level, as introduced in the previous subsection. Finally, the DH level of each new journal was computed by majority agreement among the labelling provided by the three authors. At this point in creating the list we came up with the need for two other classes: the first one is *Mega-Journal*, which is important to semantically differentiate these journals from marginally journals even if the output of this work does not change. The other is a residual category named *Not DH* that has been used only to prevent these journals for being included in the list, and it is not used any further in this work.

## 3.3. Step 3: merging DH journals from additional lists

The list of DH journals resulting from step 2 was further complemented with the Digital Humanities Literacy Guidebook (Weingart et al., 2021). We integrated it in our list as soon as we acknowledged its release. The other lists mentioned above have not been added because the journals they include were already contained in our list. The related DH level of the new journals in (Weingart et al., 2021) added to the list was also computed by the authors using majority agreement.

At the end of the third step, we obtained a list of DH journals which included 143 distinct DH journals: 34 journals labelled at DH level *Exclusively*, 18 as *Significantly*, 87 as *Marginally*, and 4 as *Mega-Journal*. The final list is made available in Zenodo (Spinaci et al., 2019) and at https://dhjournals.github.io/list.

# 4. Mapping DH research

The main aim of our study was to analyse what is the coverage of DH publications in well-known open and proprietary databases. We wanted to analyse how the DH journals in our list are represented in Crossref, Dimensions, Microsoft Academic Graph, Scopus, and Web of Science, and to show the relations (in terms of citation links) between DH publications and other SSH and non-SSH disciplines. The following sections introduce the results addressing these two aspects. As described in Section 6, all the outcomes (data, software, visualisations) are available online.

## 4.1 Coverage of DH publications in bibliographic databases

We used print and online ISSNs available in our DH journal list to retrieve all publications for each DH journal labelled *Exclusively* and *Significantly* in every citation index being considered. We decided to perform our analysis considering a specific period, i.e., between 1 January 2000 and 31 December 2018 since the databases indexed fewer DH publications before 2000 when the field also went by a different name. In addition, at the time of our study, we did not have the full availability of bibliographic and citation data in all the databases after 2018. The data of Dimensions (version dated June 2019), Web of Science (dated 13th week of 2019) and Scopus (dated April 2020) have been provided from the CWTS. The data of Microsoft Academic Graph have been downloaded from the version dated 23 January 2020 available at the Internet Archive (https://archive.org/details/mag-2020-01-23), while the data of Crossref have been obtained from the dump dated July 2020, provided by OpenCitations (Peroni et al., 2020).

| Database | Overall publications | DH publications | | |
|---|---|---|---|---|
| | | *Exclusively* | *Significantly* | *Total* |
| Crossref | 43,092,889 | 4,690 | 3,706 | 8,396 |
| Dimensions | 35,456,451 | 2,765 | 3,950 | 6,715 |
| Microsoft Academic Graph | 30,051,825 | 2,581 | 4,242 | 6,823 |
| Scopus | 28,098,657 | 2,106 | 2,802 | 4,908 |
| Web of Science | 22,119,154 | 1,472 | 1,327 | 2,799 |

**Table 1**. The number of journal articles published between 1 January 2000 and 31 December 2018, including the coverage of DH publications (only if labelled Extremely and Significantly), per database.

| Database | Exclusively | | Significantly | |
| --- | --- | --- | --- | --- |
| | *1st* | *2nd* | *1st* | *2nd* |
| Crossref | LLC (1,944) | CHum (1,555) | AI&S (1,468) | CL (951) |
| Dimensions | RIDE (797) | LLC (653) | AI&S (1,000) | CL (728) |
| Microsoft Academic Graph | LLC (602) | RIDE (416) | D-Lib (946) | CL (914) |
| Scopus | CHum (1,123) | LLC (550) | LRE (994) | AI&S (860) |
| Web of Science | RIDE (529) | LLC (339) | CL (673) | LRE (420) |

**Table 2.** The top two DH journals by the number of publications for the DH levels Exclusively and Significantly. The acronyms in the table stand for: LCC is *Literary and Linguistic Computing*, RIDE is *RIDE, a review journal for digital editions and resources*, CHum is *Computers and the Humanities*, CL is *Computational Linguistics*, AI&S is *AI & Society*, LRE is *Language Resource and Evaluation*, and D-Lib is *D-Lib Magazine*.

Tab. 1 shows the number of overall publications included in the databases. Crossref is the biggest database since it included more than 43 million publications, followed by Dimensions with more than 35 million publications, Microsoft Academic Graph with more than 30 million publications, Scopus with more than 28 million publications, and Web of Science with more than 22 million publications. Crossref was the database containing the highest number of DH publications, followed by Microsoft Academic Graph and Dimensions with a similar amount of DH publications, Scopus and, in the end, Web of Science with a relatively small amount of DH publications.

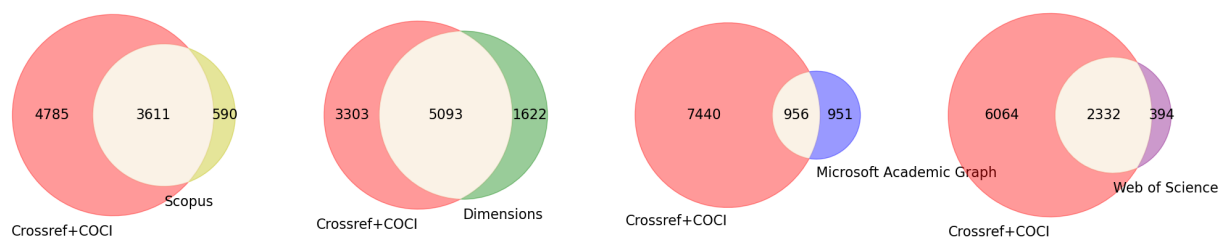

**Figure 1.** The intersection of DH publications between Crossref and all other databases based on DOIs. The other Venn diagrams are available online as detailed in Section 6.

We also analysed the overlap of DH publications across all the databases, as shown in Fig. 1. We used the DOIs to identify the same publications across databases. Indeed, Crossref uses DOIs as internal identifiers and Dimensions historically maintains DOIs along with its custom identifier. However, the other databases included many publications without any DOI assigned. Microsoft Academic Graph contained only 1,907 (out of 6,823) publications with a DOI, Scopus contained only 4,201 publications (out of 4,908) with a DOI, and in Web of Science only 2,726 publications (out of 2,799) had a DOI. As shown in Fig. 1, There is a substantial overlap between Dimensions and Crossref, while the small overlap with Microsoft Academic Graph could be explained by the small number of DOIs indexed in it.

As shown in Tab. 2, the journal *Literary and Linguistic Computing* (LCC) is one of the best represented DH journals labelled Exclusively, since it was always included among the top two journals considering the number of publications in all the databases. LCC is followed by *RIDE, a review journal for digital editions and resources*, included in the top two of three out of five databases, and *Computers and the Humanities* (CHum), included in the top two in two out of five databases. For what concern the DH journals labelled Significantly, *Computational Linguistics* (CL) was listed in the top two in all the databases but Scopus, followed by *AI & Society* (AI&S), *Language Resource and Evaluation* (LRE), and *D-Lib Magazine* (D-Lib), being listed in the top two of three, two, and one database(s), respectively.

## 4.2 Relationships between DH and other disciplines

To understand the relationship between DH publications and SSH and non-SSH fields, we analysed the citation links in all the databases introduced in Section 4.1, i.e., Dimensions, Microsoft Academic Graph, Scopus, Web of Science (WoS), and Crossref which we complemented with COCI (Heibi et al., 2019), the OpenCitations Index of Crossref open DOI-to-DOI citations (http://opencitations.net/index/coci), using the July 2020 release (OpenCitations, 2020). As before, we considered only the journal articles published from 1 January 2000 to 31 December 2018.

### 4.2.1 Retrieving publications in Arts and Humanities and Social Sciences

In Section 3 we have presented the method we have used to identify DH journals labelled Exclusively and Significantly and, consequently, DH publications contained in them. We used a similar approach to identify the SSH publications by leveraging the data in ERIH PLUS (https://dbh.nsd.uib.no/publiseringskanaler/erihplus), an academic journal index that contained (as of 30 September 2020) around 9,000 Arts & Humanities and Social Sciences journals organised in 30 disciplines – Archaeology, Art and Art History, Classical Studies, Cultural Studies, History, Library and Information Science, Linguistics, Literature, and more.

First, we classified all the 30 ERIH PLUS disciplines manually as Social Sciences (SS), Arts and Humanities (AH), or both, internally discussing and voting the best fitting domain (the classification output is available online). We then used such a new classification to label all the journals in ERIH PLUS accordingly. We labelled each journal as AH if it fell under at least one AH discipline. Otherwise, it was labelled as SS. In the labelling activities, we excluded journals already classified as DH, as described in Section 3, and *Mega-Journals* such as Nature and Science, which were not considered due to their multidisciplinary nature, making it impossible to label their articles via a journal-level category. It is worth mentioning that each journal in ERIH PLUS might specify more than one discipline. Our approach was to consider all the articles published in a DH journal as DH publications and all the other articles that were published in a journal listed in ERIH PLUS as belonging to all its ERIH PLUS disciplines. In the latter case, we used fractional counting to reach a balanced association of each publication to the ERIH PLUS disciplines assigned to the related journal. For example, if publication A belongs to two disciplines, e.g., History and Archaeology, then publication A will contribute to History and

Archaeology with an equal weight of 0.5. In this way, we weighed the contribution of each publication to different disciplines.

Next, for each database, we retrieved all publications that either were citing or were cited by DH publications. These publications were then grouped into four sets, according to the labelling the related journals had associated: *AH*, *SS*, *DH*, and *Other* (residual category). Considering these four sets, we were able to compute the distribution of outgoing and incoming citations from/to DH publications. It is worth mentioning that, in the various databases, many DH publications had no incoming and outgoing citations, as shown in Tab. 3. For example, even with similar numbers to Dimensions, most DH publications in Microsoft Academic Graph did not have citations at all, and only a few hundred publications had at most two incoming and outgoing citations.

| Database | Overall number of DH publications | DH publications without incoming citations | DH publications without outgoing citations |
| --- | --- | --- | --- |
| Crossref+COCI | 8,486 | 3,788 (44.64%) | 3,840 (45.25%) |
| Dimensions | 6,715 | 2,487 (37.04%) | 2,329 (34.68%) |
| Microsoft Academic Graph | 6,823 | 4,921 (72.12%) | 5,696 (83.48%) |
| Scopus | 4,908 | 1,110 (22.62%) | 989 (20.15%) |
| Web of Science | 2,799 | 1,059 (37.83%) | 691 (24.69%) |

**Table 3.** The number of DH publications without incoming and outgoing citations, per database.

Finally, in the analysis of such citations when considering Web of Science, we were also able to explore the publications labelled as *Other* since we had available Web of Science Research Areas (https://images.webofknowledge.com/images/help/WOS/hp_research_areas_easca.html), which are assigned to each publication which covered additional disciplines, such as Computer Science.

### 4.2.2. Spotting the relationships with DH research

In Fig. 2, we present the comparison between the citation data retrieved in all the databases. The data gathered shows that about 75-80% of publications citing or cited by DH publications belong to the set *Other*. The incoming citations from *Other* publications in Web of Science is the lowest (60%), while Dimensions and Crossref+COCI showed very similar patterns for incoming citations, and Microsoft Academic Graph is the database with the largest percentage of incoming and outgoing citations from the set *Other*. In the case of Crossref+COCI and Microsoft Academic Graph, the percentages of outgoing citations are higher than the incoming, while it is the opposite for all the other databases. Moreover, except for Microsoft Academic Graph, the absolute percentages of *AH* disciplines are higher (or very close in outgoing Crossref+COCI) than *DH*.

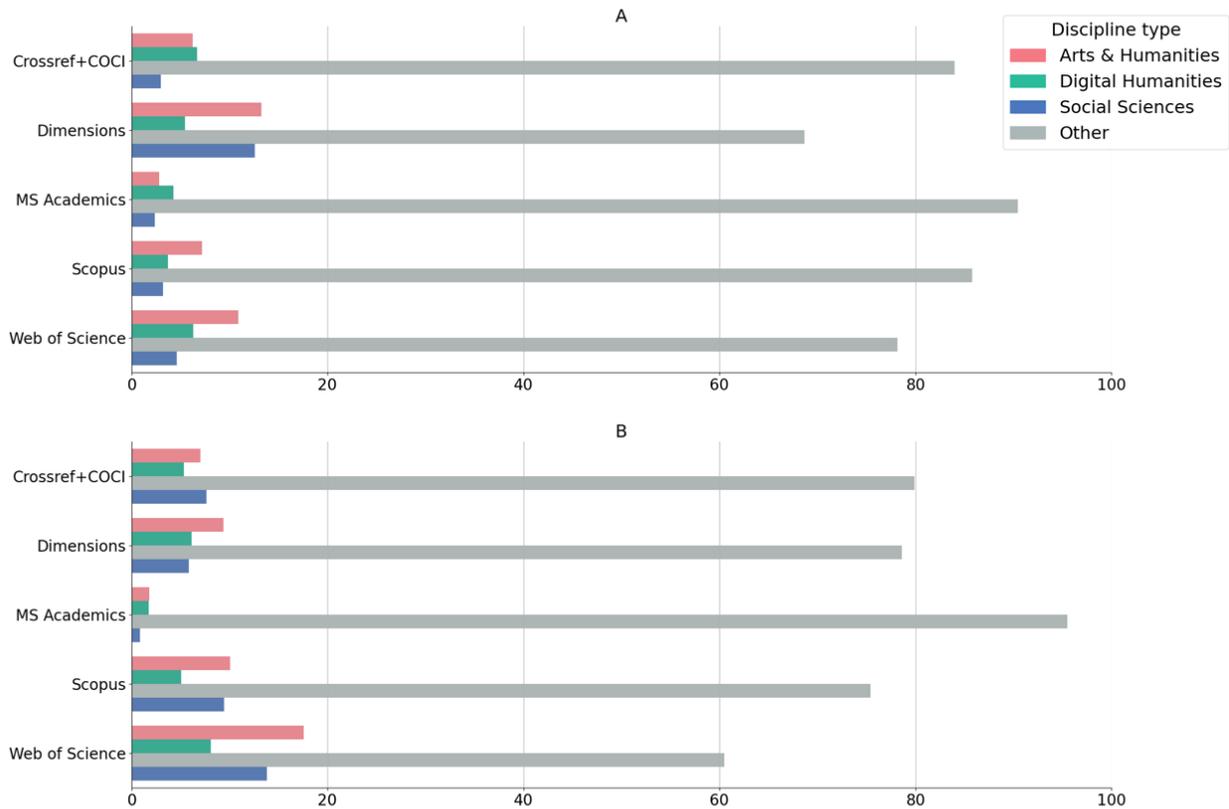

**Figure 2.** Distribution of outgoing citations from DH publications (A, top) and incoming citations to DH publications (B, bottom) for each database.

| Database | Other | DH | Linguistics | Psychology | Pedagogical & Educational Research |
|---|---|---|---|---|---|
| Crossref+COCI | 84.03 | 6.69 | 4.07 | 1.23 | 1.08 |
| Dimensions | 68.68 | 5.47 | 5.84 | 6.91 | 2.15 |
| Microsoft Academic Graph | 90.49 | 4.30 | 1.78 | 1.82 | 0.37 |
| Scopus | 85.81 | 3.73 | 4.91 | 1.43 | 0.86 |
| Web of Science | 78.18 | 6.29 | 7.58 | 2.64 | 1.01 |

**Table 4.** Distribution of outgoing citations considering main categories for each database.

| Database | Other | DH | Linguistics | Psychology | Pedagogical & Educational Research |
|---|---|---|---|---|---|
| Crossref+COCI | 79.9 | 5.35 | 3.13 | 4.15 | 1.38 |
| Dimensions | 78.6 | 6.15 | 5.59 | 2.63 | 1.75 |
| Microsoft Academic Graph | 95.5 | 1.78 | 1.30 | 0.36 | 1.12 |
| Scopus | 75.5 | 5.07 | 5.02 | 4.77 | 1.89 |
| Web of Science | 60.49 | 8.09 | 10.19 | 7.77 | 3.20 |

**Table 5.** Distribution of incoming citations considering main categories for each database.

To have a more fine-grained exploration of such data, we decoupled *AH* and *SS* publications using all their related ERIH PLUS disciplines. As anticipated in Section 4.2.1, we used fractional counting to avoid counting the same publication more than once. We show the results for all databases against the categories with higher distribution (Tab. 4-5), with a particular focus on Web of Science (Fig. 3 and 4), while the plots of remaining databases are available online, as detailed in Section 6. We decided to discuss only Crossref+COCI due to its great coverage of DH journals, and Web of Science because of the possibility of decoupling the set *Other* using the research area such as a database specified for each article, that was available at the time of the analysis.

Looking at the data for Crossref+COCI, we noticed citations from/to DH publications pointed to *Digital Humanities, Linguistics*, *Psychology*, *Pedagogical & Educational Research*, which were responsible for more than 13% of outgoing citations and 14% of incoming citations. Instead, looking at the plot drawn for Web of Science (Fig. 3), we noticed citations from/to DH publications pointed to *Digital Humanities*, *Linguistics*, *Psychology*, *Pedagogical & Educational Research*, and *Literature*, which were responsible for more than 18% of outgoing citations and 31% of incoming citations. Thus, there is a good overlap among the disciplines responsible for the main part of incoming and outgoing citations (excluding *Other*) in Crossref+COCI and Web of Science. Indeed, *Digital Humanities*, *Linguistics*, *Psychology*, and *Pedagogical & Educational Research*, are the four main citing and cited disciplines in both databases.

We unpacked the category *Other* from Fig. 3, considering the research areas specified in Web of Science (as shown in Fig. 4) and consistently using fractional counting. We stacked two coloured bars when Web of Science research areas and ERIH PLUS disciplines had the same name. In the diagram in Fig. 4, it is worth mentioning that three Computer Science (CS) research areas, namely *Artificial Intelligence*, *Information Systems*, and *Theory & Methods*, were responsible for 35.8% of outgoing citations and 20.6% of incoming citations. Computer Science disciplines take up the largest number of citations from/to DH coming from the bar *Other* shown in Fig. 3.

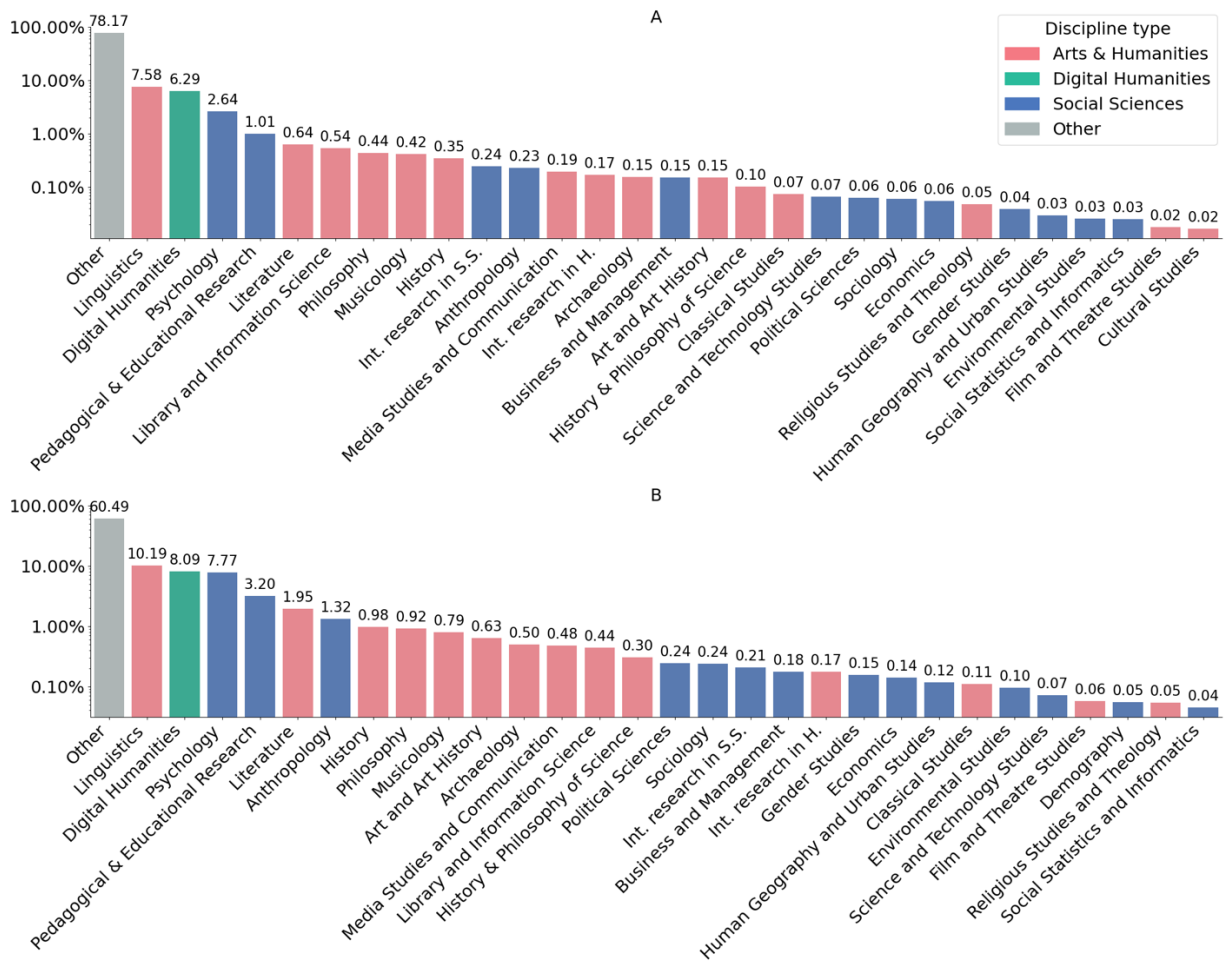

**Figure 3.** Distribution of outgoing citations (A, top) and incoming citations (B, bottom) in Web of Science. The percentages (Y-axis) are displayed using a logarithmic scale from 0 to 100%.

Finally, we also created a visualisation (shown in Fig. 5) for analysing the difference between outgoing and incoming citations from/to DH publications in Web of Science. For each category shown in Fig. 4, we calculated the difference between the outgoing and incoming citations percentages. Then, we set the percentage of incoming citations as being negative. Positive values on the x-axis mean that DH cited more the related discipline in the y-axis than the number of citation DH publications received by the same discipline. Instead, negative values on the y-axis mean that DH has been cited by the related discipline in the y-axis more than the citations DH did to the same discipline. The results show that the difference is positive, mainly in the case of Computer Science Research Areas, meaning that DH cited Computer Science more than the number of citations it received from the same discipline. Instead, we found that Digital Humanities and the ERIH-PLUS disciplines Psychology and Pedagogy had an opposite behaviour – they cited DH more than the number of citations they received from DH publications. Linguistics is an interesting case because its citation behaviour was different if the category comes from ERIH PLUS (coloured in red) or WoS Subject Categories (coloured in grey). For this reason, the plot shows two bars with different colours. We leave the further exploration of this finding to future studies.

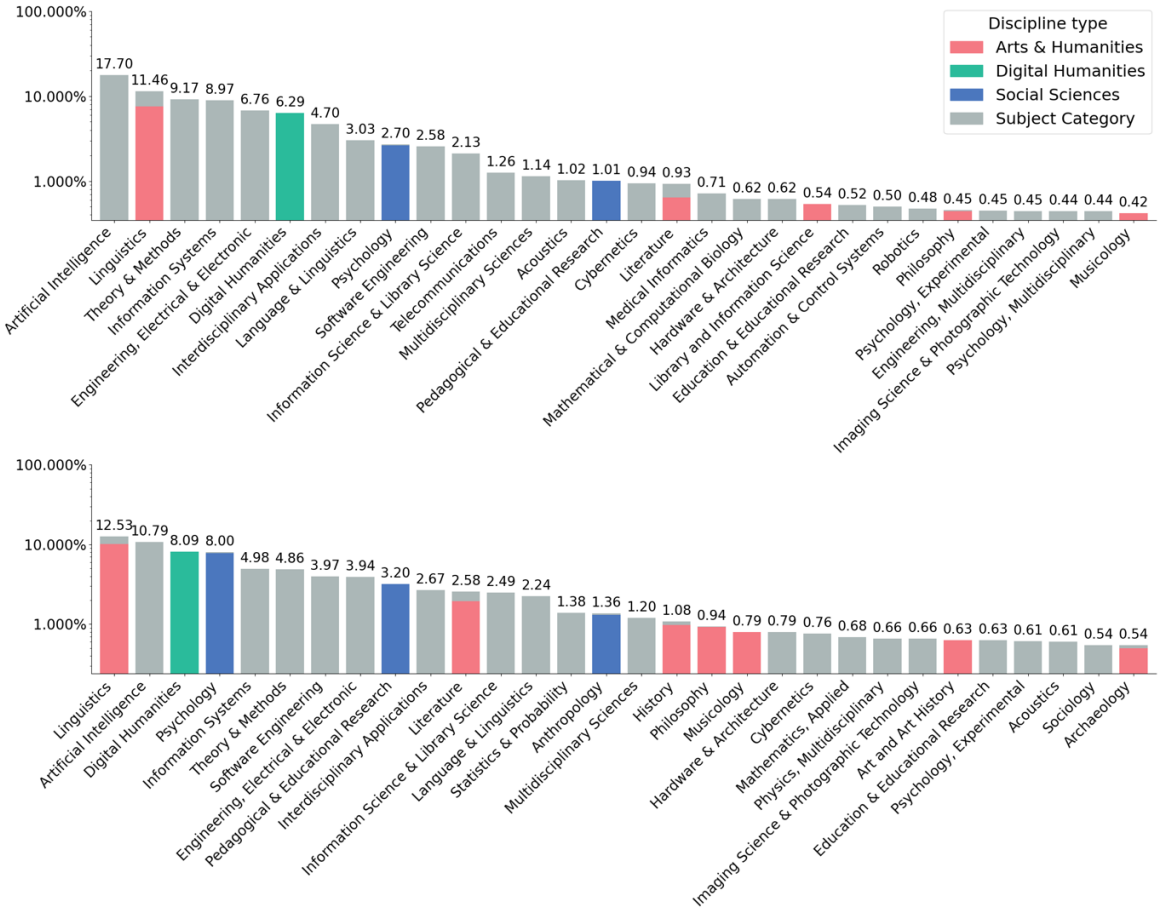

**Figure 4.** Distribution of outgoing citations (top) and incoming citations (bottom) in Web of Science, expanding the *Other* bar in Fig. 3 with the research areas defined in Web of Science. The percentages (Y-axis) are displayed using a logarithmic scale from 0 to 100%.

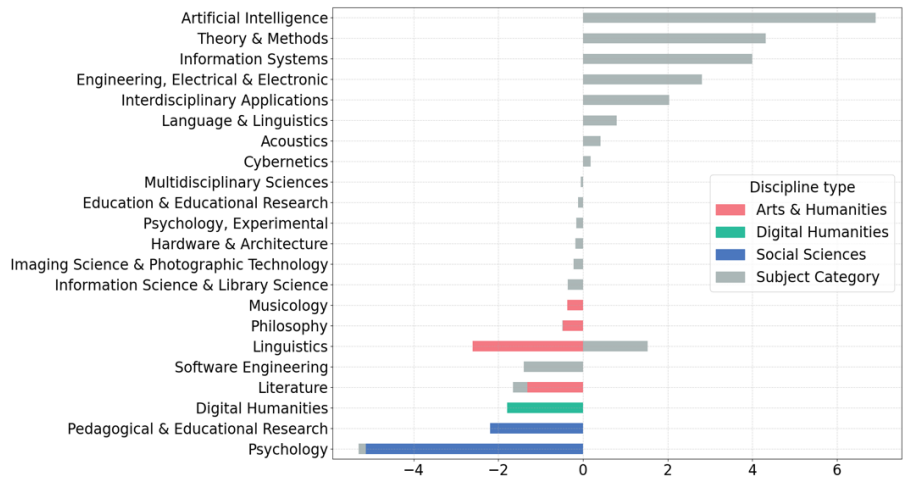

**Figure 5.** Difference between outgoing and incoming citations from/to DH publications shown considering the disciplines in Fig. 4.

# 5. Discussion and conclusions

This work aimed to identify Digital Humanities (DH) publications across bibliographic data sources and understand their coverage and the citation relationship between DH publications, the Humanities, Social Sciences, and other disciplines. The methodology we proposed involved the creation of a list of DH journals (described in Section 3) to use as an external resource to identify DH publications in open and proprietary bibliographic databases.

The bibliographic databases we used in our analysis were Crossref complemented with the citation data in COCI, Dimensions, Microsoft Academic Graph, Scopus, and Web of Science. We found that Crossref is the database with the highest coverage of DH publications, followed in order of decreasing coverage by Microsoft Academic Graph, Dimensions, Scopus and Web of Science. The combined use of Crossref and Dimension can be beneficial in terms of coverage since these databases are in part complementary and together include more than 10,000 unique DH publications (Fig. 1). Importantly, the bibliographic data in Crossref complemented with citation data from COCI are openly available. ERIH PLUS played a crucial part in identifying Arts & Humanities (AH) and Social Sciences (SS) publications, even if it does not cover all the SSH journals published in the world. Therefore, as shown in Fig. 2, the residual category *Other* was responsible for 80% of such incoming and outgoing DH citations in all databases except Microsoft Academic Graph (90-95%) and Web of Science (58-64%).

DH publications are connected via citations primarily to research in Computer Science, Linguistics, Psychology, and Pedagogical & Educational Research (Tab. 4-5 and Fig. 3-5). Our results confirm in this respect the outcomes of previous works (Nyhan et al., 2014; Scholes et al., 2008; Sula et al., 2019). In addition, from our results it appears that DH publications might act as a bridge between SSH disciplines – e.g., Linguistics, Psychology, and Pedagogical & Educational Research, that cite DH research – and Computer Science – that is cited mainly by DH research (Fig. 5). Furthermore, it is worth mentioning that Computer Science is also responsible for a sizable share of citations given to DH research. We speculate that DH research might act as a hub for Computer Science methods to diffuse in SSH disciplines, yet this warrants further study. Computer Science and DH certainly show reciprocal interest (Fig. 5). We might again speculate and suggest that computer scientists perceive DH as a concrete domain for the application of computational theories and methods. It is equally possible that strong bounds between DH and other Humanities disciplines also exist. However, they may be manifested by citations to and from books (which are not included in our analysis) or otherwise elude the coverage of existing publication databases.

This project also presents some limitations. The list of DH journals we created might not fully represents DH research published in journals. For example, it did not consider books and other research published in DH conferences and workshops proceedings. In addition, a bias could have been introduced in the list of DH journals. Indeed, we involved human experts in the first step of its creation (Section 3.1), and this could have introduced experts' subjective perception of the DH domain within the list. A further limitation could be related to the use of Web of Science during the second step of the methodology to update the list of DH journals (Section 3.2). As shown in a recent study (Martín-Martín et al., 2020), the coverage of Humanities journals in Web of Science is far from complete, which could have affected the identification of DH journals that might be missed in Web of Science. As such, the list of DH journals we created provided a partial, yet relevant, view of a complex domain, and future studies may consider addressing all these limitations to improve the list described in this article. Finally, it is worth mentioning that the choice of the specific temporal window 2000-2018 in our experiment could have produced some bias in the results. Assuming that the canonical beginning of DH is dated around 1960 and considering that some DH journals such as *Computers and the Humanities* and *Literary and Linguistic Computing* were created in 1966 and 1986, respectively, many DH publications could have been cut off from the analysis. The use of a broader time coverage will, though, be investigated in future studies.

# 6. Data and code availability

In this article, we presented various data in visual forms. In most of the cases, we decided to expose only the data relative to Crossref, COCI and Web of Science. However, the discussion and the results are also extended to the other databases, i.e., Scopus, Dimensions and Microsoft Academic Graph. All the pictures and tables of all the databases are available at https://github.com/dhjournals/code/tree/master/output, while the list of DH journals we created is available on Zenodo (Spinaci et al., 2019). Moreover, we released all code and data used for this study on Github at https://github.com/dhjournals/code. It is worth mentioning that only the scripts for analysing Crossref+COCI and Microsoft Academic Graph are available because such datasets are the only ones available in Open Access.

# 7. Acknowledgements

The authors would like to thank the Centre for Science and Technology Studies (CWTS), Leiden University, for providing access to their databases and computing facilities. This work was in part conducted when GS was visiting CWTS. The work of Silvio Peroni has been partially funded by the European Union's Horizon 2020 research and innovation program under grant agreement No 101017452 (OpenAIRE-Nexus).